\documentclass[9pt,conference]{IEEEtran}
\usepackage{dcase2026}
\usepackage{acronym}
\acrodef{ae}[AE]{autoencoder}
\acrodef{auc}[AUC]{area under the \ac{roc} curve}
\acrodef{asd}[ASD]{anomalous sound detection}
\acrodef{knn}[kNN]{$k$-nearest neighbor}
\acrodef{lora}[LoRA]{low-rank adaptation}
\acrodef{mlp}[MLP]{multilayer perceptron}
\acrodef{pauc}[pAUC]{partial \ac{auc}}
\acrodef{roc}[ROC]{receiver operating characteristic}
\acrodef{ssl}[SSL]{self-supervised learning}
\acrodef{sota}[SOTA]{state-of-the-art}
\acrodef{asr}[ASR]{automatic speech recognition}
\acrodef{tse}[TSE]{target sound extraction}
\acrodef{mse}[MSE]{mean squared error}
\acrodef{mhsa}[MHSA]{multi-head self-attention}
\acrodef{swiglu}[SwiGLU]{Swish gated linear unit}
\acrodef{ffn}[FFN]{feed-forward network}
\acrodef{sid}[SID]{speaker identification}
\acrodef{er}[ER]{emotion recognition}
\acrodef{map}[mAP]{mean average precision}
\acrodef{snr}[SNR]{signal-to-noise ratio}
\acrodef{dbeats}[DBEATs]{denoising BEATs}
\acrodef{nabeats}[NABEATs]{noise-aware BEATs}
\acrodef{ss}[SS]{spectral subtraction}
\acrodef{na}[NA]{noise-aware}
\acrodef{ca}[CA]{cross-attention}
\acrodef{rdp}[RDP]{relative deviation pooling}
\acrodef{ema}[EMA]{exponential moving average}
\acrodef{naasd}[NA-ASD]{noise-aware \ac{asd}}
\acrodef{nassl}[NA-SSL]{noise-aware \ac{ssl}}
\usepackage{tikz}
\usetikzlibrary{arrows.meta,positioning,fit,calc}
\usepackage{booktabs}
\usepackage{makecell}
\usepackage{tabularx}
\usepackage{multirow}
\usepackage[outline]{contour}
\usepackage{adjustbox}

\usepackage{bm} %

\usepackage{amsmath,graphicx,url,times}

\title{Anomalous Sound Detection Meets Noise-Aware Self-Supervised Learning} %

\name{{\shortstack[c]{Takuya Fujimura$^{1,2*}$\thanks{*This work was done during an internship at MERL.}, Gordon Wichern$^{1}$, Yoshiki Masuyama$^{1}$, Christoph Boeddeker$^{1}$, \\ Kohei Saijo$^{3}$, Julius Richter$^{1}$, Takahiro Edo$^{1}$, Jonathan Le Roux$^{1}$}}}
\address{$^1$Mitsubishi Electric Research Laboratories (MERL), Cambridge, USA, $^2$Nagoya University, Nagoya, Japan,\\$^3$Information Technology R\&D Center, Mitsubishi Electric Corporation, Kanagawa, Japan}

\begin{document}
\bstctlcite{BSTcontrol}
\maketitle

\begin{abstract}
In this paper, we introduce noise-aware self-supervised learning (NA-SSL) models for noise-aware anomalous sound detection (NA-ASD).
NA-ASD is an ASD task with two-channel audio recordings, where one microphone is located close to the target machine and the other is located farther away to capture noise.
For this task, we simulate two-channel recordings using diverse audio datasets and train NA-SSL models to extract clean SSL representations of the close-microphone signal by using the far-microphone recording dominated by background noise as auxiliary information.
The NA-SSL models are then used as frontends in the standard ASD framework.
Our experimental evaluation on the DCASE 2026 Challenge Task 2 development dataset demonstrates the effectiveness of the NA-SSL framework across three base SSL models (BEATs, EAT, and Dasheng), both with and without discriminative fine-tuning.
Furthermore, the challenge results proved the effectiveness of the proposed approach, where the NA-BEATs system won the challenge by a large margin, achieving an official score of 70.24\%, while the second-place system achieved 65.46\%.
\end{abstract}


\section{Introduction}
\label{sec:intro}

\Ac{asd} aims to detect mechanical failures from machine sounds~\cite{koizumi2020description,kawaguchi2021description,dohi2022description,dohi2023description,nishida2024description,nishida2025description,nishida2026description,fujimura2025asdkit}.
Since it is infeasible to exhaustively collect rare and diverse anomalous sounds, \ac{asd} systems are developed using only normal sounds.
One major challenge in \ac{asd} is the presence of severe background noise in factory environments during both training data collection and testing.
Motivated by this challenge, \ac{naasd} considers the use of auxiliary noise information~\cite{nishida2026description}.
\Ac{naasd} employs a practical two-microphone recording setup, where one microphone is placed close to the target machine and the other is placed farther away.
Although the far-microphone signal can also contain the target machine sound, the target sound is less dominant than in the close-microphone signal, making the far-microphone signal a useful auxiliary source of noise information.
The \ac{naasd} task was newly introduced in DCASE 2026 Challenge Task~2~\cite{nishida2026description}, and there is strong interest in developing \ac{asd} systems that can effectively utilize such auxiliary noise information.

A recent trend in standard \ac{asd} tasks is the use of audio \ac{ssl} models~\cite{saengthong2025deep,saengthong2026sub,wilkinghoff2026temporal,jiang2024anopatch,han2025exploring,fujimura2025asdkit}. %
\Ac{ssl} models effectively capture subtle differences between normal and anomalous sounds, enabling training-free \ac{asd} by simply computing distances between a test sample and normal training samples in the representation space~\cite{saengthong2025deep,saengthong2026sub,wilkinghoff2026temporal}.
\Ac{ssl} models fine-tuned on machine sound datasets through classification tasks using machine-information labels capture more machine-specific information and achieve \acl{sota} performance~\cite{jiang2024anopatch,han2025exploring,fujimura2025asdkit}.
However, under noisy conditions, such representations essentially encode irrelevant background noise as well, making it difficult to focus on target machine sounds.
One possible solution is denoising training specialized for extracting specific target machine sounds, as in WavLM~\cite{chen2022wavlm}.
However, it is not applicable to \ac{asd} tasks because clean target machine sounds are unavailable as training data.

A promising approach to handling noisy conditions in audio \ac{ssl} models is the \ac{nassl} framework~\cite{fujimura2026nabeats}.
\Ac{nassl} models are pre-trained on diverse audio signals to extract clean representations from noisy signals using auxiliary noise inputs.
Because \ac{nassl} models perform conditional denoising based on the provided auxiliary noise information, they do not necessarily require clean target sounds or noise samples specific to the downstream task as pre-training data.
That is, no clean machine sounds are used for NA-SSL training, only clean sounds from external datasets such as FSD50K~\cite{fonseca2022FSD50K}.
In~\cite{fujimura2026nabeats}, one specific implementation of the \ac{nassl} framework was investigated, in which NA-BEATs was built on BEATs~\cite{chen2023beats} and noise-only segments extracted from noisy input signals were used as auxiliary noise inputs.
More generally, the framework is expected to accommodate any auxiliary signal that captures noise information, such as the far-microphone signal available in the \ac{naasd} task.

\begin{figure}[t]
    \centering
    \input{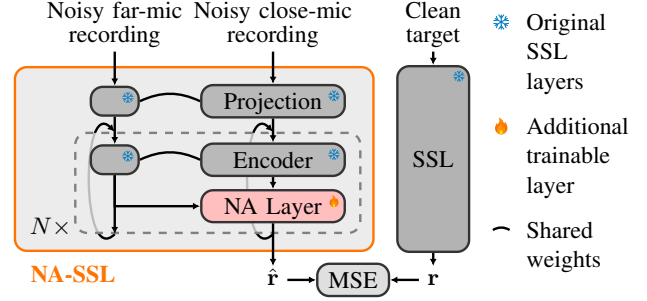}
    \vspace{-5pt}
    \caption{
        Overview of NA-SSL, adapted from \cite{fujimura2026nabeats}. The model estimates clean SSL representations $\mathbf r$ from two-channel noisy signals, with one microphone placed close to the target sound source and the other placed farther away to capture background noise information.
    }
    \vspace{-11pt}
    \label{fig:nassl}
\end{figure}

In this paper, we explore the use of \ac{nassl} models for the \ac{naasd} task.
To this end, we simulate two-channel recordings using diverse audio signals and pre-train \ac{nassl} models to extract clean representations from two-channel noisy signals, as shown in \cref{fig:nassl}.
The \ac{nassl} models learn conditional denoising by exploiting the difference between the two channels: target sound information is more prominent in the close-microphone signal than in the far-microphone signal, whereas noise information is more prominent in the far-microphone signal.
Through this conditional denoising scheme, \ac{nassl} models can provide denoised representations for the downstream \ac{asd} task without requiring clean target machine sounds as pre-training data.
Furthermore, in addition to NA-BEATs, we newly design NA extensions of EAT~\cite{chen2024eat} and Dasheng~\cite{dinkel2024scaling}.
Experimental evaluations on the DCASE 2026 Challenge Task~2 development dataset show that \ac{nassl} substantially improves performance across the three base \ac{ssl} models, both with and without discriminative fine-tuning.
In the official challenge evaluation, NA-BEATs with discriminative fine-tuning achieved the best performance among 175 system submissions, with an official score of 70.24\%, whereas the second-place system achieved 65.46\% and the official baseline system achieved 59.80\%.

\section{SSL-based ASD}
\label{sec:ssl_asd}
Typical \ac{asd} systems consist of a \textit{frontend} and a \textit{backend}~\cite{fujimura2025asdkit}.
The frontend extracts representations from machine sounds, and the backend computes anomaly scores based on distances in the representation space.
Recent studies have extensively investigated audio \ac{ssl} models as frontends~\cite{saengthong2025deep,saengthong2026sub,wilkinghoff2026temporal,jiang2024anopatch,han2025exploring,fujimura2025asdkit}, along with advances in backend techniques~\cite{saengthong2025deep,saengthong2026sub,wilkinghoff2026temporal}. %

\subsection{SSL models}
\textbf{BEATs}~\cite{chen2023beats} is an audio \ac{ssl} model widely adopted for \ac{asd}~\cite{saengthong2025deep,saengthong2026sub,wilkinghoff2026temporal,jiang2024anopatch,han2025exploring,fujimura2025asdkit}. %
It performs masked-prediction-based \ac{ssl} in a discrete token space.
A key feature of BEATs is its iterative training pipeline, in which the acoustic tokenizer and the \ac{ssl} model are alternately refined through knowledge distillation.
\textbf{EAT}~\cite{chen2024eat} has also demonstrated strong \ac{asd} performance~\cite{saengthong2026sub,wilkinghoff2026temporal,fujimura2025asdkit}. %
It is based on a masked bootstrapping framework, in which a student model is trained to reconstruct the latent representations of a teacher model at masked positions.
It also introduces utterance-level losses and a CLS token to capture global information.
\textbf{Dasheng}~\cite{dinkel2024scaling} is a recent \ac{ssl} model based on a masked autoencoder framework and is trained on 272k hours of diverse audio data.
Dasheng takes mel-spectrogram segments as input, whereas BEATs and EAT take mel-spectrogram patches as input.
These models convert the input sequence into a representation sequence $\mathbf{r}\in\mathbb{R}^{L\times D}$, where $L$ denotes the number of mel-spectrogram patches or segments and $D$ denotes the representation dimension.
The representation is then either directly used in the backend or further fine-tuned on machine sound datasets.

\subsection{Discriminative fine-tuning}
Fine-tuning is performed based on a classification task using machine-information labels~\cite{jiang2024anopatch,han2025exploring}.
In recent DCASE datasets~\cite{nishida2024description,nishida2025description,nishida2026description}, the machine-information labels in the training data include the machine type (e.g., bearing), domain (source or target), and machine-dependent fine-grained attribute labels (e.g., rotation speed).
Here, different attributes are included in the source and target domains.
Also, some machine types do not have attribute labels.
For such machine types, pseudo-labeling is employed~\cite{fujimura2025improvements,han2025exploring}.
Pseudo attribute labels are generated by applying clustering to the representations of the training data.
The pseudo attribute labels are then combined with the other available coarse labels of the target machine (i.e., the machine-type and domain labels) and the attribute labels of the other machines to form a multi-class classification task.

\subsection{Backend techniques}
\label{sec:ssl_asd_backend}
In the backend, the \ac{ssl} representation sequence $\mathbf{r}$ is aggregated to calculate the anomaly score.
Saengthong et al.~\cite{saengthong2025deep} proposed an aggregation technique that utilizes frequency information.
This technique reshapes $\mathbf{r}\in\mathbb{R}^{L\times D}$ into $\mathbf{r}_{\rm TF}\in\mathbb{R}^{T\times F\times D}$, where $T$ and $F$ denote the numbers of mel-spectrogram patches along the time and frequency axes, respectively.
It then applies average pooling along the time axis and concatenates the frequency axis to obtain a single $FD$-dimensional frequency-preserving representation.
This representation was shown to improve backend performance compared with a global representation averaged over both the time and frequency axes~\cite{saengthong2025deep}.

BEAM~\cite{saengthong2026sub} is another technique that builds on a frequency-preserving representation.
It constructs a $D$-dimensional representation memory bank for each frequency index.
An anomaly score is calculated separately for each frequency index and then averaged across the frequency indices to obtain a sample-level anomaly score.
This technique mitigates undesirably high anomaly scores caused by variations within normal sounds.
For example, an unseen combination of machine sound and noise can yield a high anomaly score, even when each frequency component has been separately observed in the training data.
By avoiding strict joint comparison across frequencies, this technique can improve robustness to normal variations while retaining sensitivity to anomalies in each frequency.

\Ac{rdp}~\cite{wilkinghoff2026temporal} is a pooling technique designed as an alternative to simple average pooling.
It computes a weighted average of the representation sequence, assigning each representation $\mathbf{r}_l$ a weight $w_l= {(1+\hat{d}_l)^\gamma}/{\sum_{l'}(1+\hat{d}_{l'})^\gamma}$ where $\hat{d}_l$ is the Euclidean distance between $\mathbf{r}_l$ and the sequence average, normalized by the maximum such distance within the sequence, and $\gamma$ is a hyperparameter.
This emphasizes representations that deviate from the average, thereby preserving temporally localized anomalies that are smoothed out by typical average pooling.

\section{Proposed method}
We introduce \ac{nassl} models to the \ac{naasd} task.
We follow the training procedure of the previous NA-BEATs implementation~\cite{fujimura2026nabeats} but modify the auxiliary noise condition to match the \ac{naasd} task.
Also, we design multiple \ac{nassl} models.
Specifically, we design NA-BEATs, NA-EAT, and NA-Dasheng in the same manner based on BEATs (\textit{BEATs\_iter3.pt}), EAT (\textit{EAT-base\_epoch30\_pretrain}), and Dasheng (\textit{dasheng-base}), respectively.

\subsection{Noise-aware self-supervised learning (NA-SSL)}
\label{sec:nassl}
Following \cite{fujimura2026nabeats}, we train an \ac{nassl} model by distillation with a \ac{mse} loss between the clean target representation sequence $\mathbf{r}\!\in\!\mathbb{R}^{L\times D}$ from the frozen original \ac{ssl} model and the estimated representation sequence $\hat{\mathbf r}\!\in\!\mathbb{R}^{L\times D}$~\cite{fujimura2026nabeats}, given by
\begin{align}
    \mathbf r\!&=\!\mathrm{SSL}(\mathbf s),\\
    \hat{\mathbf r}\!&=\!\mathrm{NASSL}({\mathbf x}, \mathbf{n}^\prime),
    \label{eq:nassl}
\end{align}
where $\mathbf{s}$, $\mathbf{x}$, and $\mathbf{n}^\prime$ denote clean target sound, noisy target sound, and auxiliary noise signal, respectively.
For the \ac{naasd} task, we use the close-microphone signal as $\mathbf{x}$ and the far-microphone signal as $\mathbf{n}^\prime$.
We define the clean target sound $\mathbf{s}$ as the signal obtained at the close microphone using a room impulse response including early reflections within 50~ms.
For NA-EAT, which produces a CLS token in addition to the representation sequence, we apply the \ac{mse} loss to both outputs with equal weights.

For all base \ac{ssl} models, we construct NA extensions by inserting trainable NA layers into the frozen original \ac{ssl} model after each Transformer layer, as in \cite{fujimura2026nabeats}.
Here, BEATs, EAT, and Dasheng all have a projection layer and 12 Transformer layers.
Both the close-microphone signal $\mathbf{x}$ and the far-microphone signal $\mathbf{n}^\prime$ are passed through the original base \ac{ssl} model.
After each original Transformer layer, an NA layer refines the noisy target representation $\mathbf{z}_\mathbf{x}\!\in\!\mathbb{R}^{L\times D}$ using the corresponding noise representation $\mathbf{z}_{\mathbf{n}^\prime}\!\in\!\mathbb{R}^{L\times D}$.
Specifically, the cross-attention-based NA layer works as follows~\cite{fujimura2026nabeats}:
\begin{align}
    \mathbf{z}_\mathbf{x} &\xleftarrow{} \mathbf{z}_\mathbf{x} + \mathrm{MHCA}(\mathrm{RMSNorm}(\mathbf{z}_\mathbf{x}), \mathrm{RMSNorm}(\mathbf{z}_{\mathbf{n}^\prime})),\\
    \mathbf{z}_\mathbf{x} &\xleftarrow{} \mathbf{z}_\mathbf{x} + \mathrm{FFN}_\mathrm{SwiGLU}(\mathrm{RMSNorm}(\mathbf{z}_\mathbf{x})),
    \label{eq:ffn}
\end{align}
where $\mathrm{MHCA}$ denotes multi-head cross-attention~\cite{vaswani2017attention}, with the first argument used as the query and the second argument used as the key and value.
$\mathrm{RMSNorm}$ denotes root mean square normalization~\cite{Zhang2019RMSNorm}, and $\mathrm{FFN}_\mathrm{SwiGLU}$ denotes a \ac{swiglu}-based \ac{ffn}~\cite{shazeer2020glu}.
The NA layer incorporates auxiliary noise information to refine the noisy representation and was shown to improve performance under noisy conditions in \cite{fujimura2026nabeats}.

\subsection{Simulation of two-channel recordings}
For training the \ac{nassl} models, we simulate two-channel recordings using diverse audio signals.
FSD50K~\cite{fonseca2022FSD50K} is used as the target-sound dataset, while WHAM!48kHz~\cite{wichern2019WHAM}, DEMAND~\cite{thiemann2013diverse}, and QUT-NOISE~\cite{dean2010qut} are used as the noise datasets.
All signals are resampled to 16~kHz and randomly cropped or zero-padded to 10~s.

\begin{figure}[t]
    \centering
    \definecolor{roommargin}{HTML}{F7A3A8}
\definecolor{roomfloor}{HTML}{FFFFFF}

\def\RoomScale{0.82}
\def\RoomLengthM{8}
\def\RoomWidthM{4}
\def\TargetMarginM{0.7}
\def\NoiseCornerMarginM{0.4}
\def\TargetCloseDistanceM{0.9}
\def\TargetFarDistanceM{1.8}

\def\TargetXFraction{0.32}
\def\TargetYFraction{0.25}
\def\FarMicAngleDeg{130}

\def\DimensionOffsetM{0.35}
\def\SourceMarginLabelExtraM{0.75}
\def\NoiseMarginLabelExtraM{0.45}
\def\MarginLabelXOffset{0.06}
\def\SourceMarginLabelYOffset{0.15}
\def\CloseDistanceLabelYOffset{0.18}
\def\FarDistanceLabelPos{0.6}
\def\FarDistanceLabelXOffset{-0.08}
\def\FarDistanceLabelYOffset{0.05}
\def\LegendInset{0.05}
\def\RoomSymbolRadius{4.5pt}

\pgfmathsetmacro{\RoomLength}{\RoomLengthM*\RoomScale}
\pgfmathsetmacro{\RoomWidth}{\RoomWidthM*\RoomScale}
\pgfmathsetmacro{\TargetMargin}{\TargetMarginM*\RoomScale}
\pgfmathsetmacro{\NoiseCornerMargin}{\NoiseCornerMarginM*\RoomScale}
\pgfmathsetmacro{\TargetCloseDistance}{\TargetCloseDistanceM*\RoomScale}
\pgfmathsetmacro{\TargetFarDistance}{\TargetFarDistanceM*\RoomScale}
\pgfmathsetmacro{\DimensionOffset}{\DimensionOffsetM*\RoomScale}
\pgfmathsetmacro{\SourceMarginLabelX}{(\TargetMarginM + \SourceMarginLabelExtraM)*\RoomScale}
\pgfmathsetmacro{\NoiseMarginLabelX}{(\NoiseCornerMarginM + \NoiseMarginLabelExtraM)*\RoomScale}

\pgfmathsetmacro{\TargetX}{(\TargetMarginM + \TargetXFraction*(\RoomLengthM - 2*\TargetMarginM))*\RoomScale}
\pgfmathsetmacro{\TargetY}{(\TargetMarginM + \TargetYFraction*(\RoomWidthM - 2*\TargetMarginM))*\RoomScale}
\pgfmathsetmacro{\CloseMicX}{\TargetX + \TargetCloseDistance}
\pgfmathsetmacro{\CloseMicY}{\TargetY}
\pgfmathsetmacro{\FarMicX}{\TargetX + \TargetFarDistance*cos(\FarMicAngleDeg)}
\pgfmathsetmacro{\FarMicY}{\TargetY + \TargetFarDistance*sin(\FarMicAngleDeg)}

\newcommand{\RoomLengthLabel}{$\sim\mathcal{U}(3.0,8.0)$~m}
\newcommand{\RoomWidthLabel}{$\sim\mathcal{U}(3.0,8.0)$~m}
\newcommand{\TargetMarginLabel}{0.5~m}
\newcommand{\NoiseCornerMarginLabel}{0.2~m}
\newcommand{\TargetCloseDistanceLabel}{0.05~m}
\newcommand{\TargetFarDistanceLabel}{$\sim\mathcal{U}(0.1,1.0)$~m}
\newcommand{\DrawNoiseSource}[3]{%
    \begin{scope}[shift={(#1,#2)}, rotate=#3]
        \draw[fill=gray!50, draw=black, line width=0.5pt]
            (-0.15,-0.07) rectangle (-0.04,0.07);
        \draw[fill=gray!35, draw=black, line width=0.5pt]
            (-0.04,-0.12) -- (0.15,-0.20) -- (0.15,0.20) -- (-0.04,0.12) -- cycle;
    \end{scope}
}
\newcommand{\DrawTargetSource}[1]{%
    \begin{scope}[shift={(#1)}]
        \draw[draw=black, fill=gray!35, line width=0.45pt]
            (0,0) circle[radius=\RoomSymbolRadius];
        \node[font=\footnotesize, inner sep=0pt] at (0,0) {T};
    \end{scope}
}
\newcommand{\DrawMicrophone}[2]{%
    \begin{scope}[shift={(#1)}]
        \draw[draw=black, fill=white, line width=0.45pt]
            (0,0) circle[radius=\RoomSymbolRadius];
        \node[font=\footnotesize, inner sep=0pt] at (0,0) {#2};
    \end{scope}
}

\begin{tikzpicture}[
    x=1cm,
    y=1cm,
    font=\footnotesize,
    >={Latex[length=1.4mm,width=1.1mm]},
    wall/.style={draw=black, line width=0.7pt},
    dim/.style={<->, draw=black, line width=0.55pt},
    note/.style={font=\footnotesize, inner sep=1pt, align=center},
    legend/.style={
        font=\footnotesize,
        inner sep=2pt,
        align=left,
        draw=black!35,
        fill=white,
        fill opacity=0.88,
        text opacity=1,
        rounded corners=1pt
    }
]

\draw[fill=roommargin, wall] (0,0) rectangle (\RoomLength,\RoomWidth);
\draw[fill=roomfloor, wall]
    (\TargetMargin,\TargetMargin)
    rectangle ({\RoomLength-\TargetMargin},{\RoomWidth-\TargetMargin});
\draw[wall]
    (\TargetMargin,\TargetMargin)
    rectangle ({\RoomLength-\TargetMargin},{\RoomWidth-\TargetMargin});
\draw[draw=black, line width=0.45pt]
    (\NoiseCornerMargin,\NoiseCornerMargin)
    rectangle ({\RoomLength-\NoiseCornerMargin},{\RoomWidth-\NoiseCornerMargin});

\draw[dim] (0,-\DimensionOffset) -- (\RoomLength,-\DimensionOffset)
    node[note, midway, below=1pt] {\RoomLengthLabel};
\draw[dim] ({\RoomLength+\DimensionOffset},0) -- ({\RoomLength+\DimensionOffset},\RoomWidth);
\node[note, rotate=90] at ({\RoomLength+1.75*\DimensionOffset},{0.5*\RoomWidth})
    {\RoomWidthLabel};

\draw[dim] (\SourceMarginLabelX,\RoomWidth) -- (\SourceMarginLabelX,{\RoomWidth-\TargetMargin});
\node[note, anchor=west] at ({\SourceMarginLabelX+\MarginLabelXOffset},{\RoomWidth-\SourceMarginLabelYOffset})
    {\TargetMarginLabel};
\draw[dim] (\NoiseMarginLabelX,0) -- (\NoiseMarginLabelX,\NoiseCornerMargin);
\node[note, anchor=west] at ({\NoiseMarginLabelX+\MarginLabelXOffset},{0.5*\NoiseCornerMargin})
    {\NoiseCornerMarginLabel};

\DrawNoiseSource{\NoiseCornerMargin}{\NoiseCornerMargin}{-135}
\DrawNoiseSource{{\RoomLength-\NoiseCornerMargin}}{\NoiseCornerMargin}{-45}
\DrawNoiseSource{\NoiseCornerMargin}{{\RoomWidth-\NoiseCornerMargin}}{135}
\DrawNoiseSource{{\RoomLength-\NoiseCornerMargin}}{{\RoomWidth-\NoiseCornerMargin}}{45}

\coordinate (target) at (\TargetX,\TargetY);
\coordinate (closemic) at (\CloseMicX,\CloseMicY);
\coordinate (farmic) at (\FarMicX,\FarMicY);

\coordinate (targetcloseedge) at ($(target)!\RoomSymbolRadius!(closemic)$);
\coordinate (closetargetedge) at ($(closemic)!\RoomSymbolRadius!(target)$);
\coordinate (targetfaredge) at ($(target)!\RoomSymbolRadius!(farmic)$);
\coordinate (fartargetedge) at ($(farmic)!\RoomSymbolRadius!(target)$);

\draw[dim] (targetcloseedge) -- (closetargetedge);
\node[note, anchor=north] at ($(targetcloseedge)!0.5!(closetargetedge)+(0.1,-\CloseDistanceLabelYOffset)$)
    {\TargetCloseDistanceLabel};
\draw[dim] (targetfaredge) -- (fartargetedge);
\node[note, anchor=south west] at ($(targetfaredge)!\FarDistanceLabelPos!(fartargetedge)+(\FarDistanceLabelXOffset,\FarDistanceLabelYOffset)$)
    {\TargetFarDistanceLabel};

\DrawTargetSource{target}
\DrawMicrophone{closemic}{C}
\DrawMicrophone{farmic}{F}

\node[legend, anchor=south east] at ({\RoomLength-\TargetMargin-\LegendInset},{\TargetMargin+\LegendInset})
    {T: target source\\C: close mic\\F: far mic\\\tikz[baseline=-0.5ex, x=0.45cm, y=0.45cm]{\DrawNoiseSource{0}{0}{0, rounded corners=0pt}}: noise sources};

\end{tikzpicture}%
    \caption{
        Room geometry for simulation.
        The room height is sampled from $\mathcal{U}(1.5, 3.0)$~m, and all sources and microphones are placed at a height of 1.0~m.
    }
    \label{fig:room_simulation}
\end{figure}
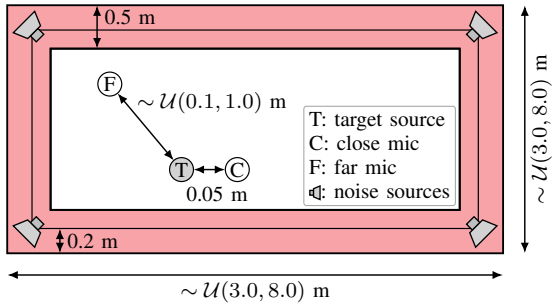

We simulate room impulse responses using Pyroomacoustics~\cite{scheibler2018pyroomacoustics}, as shown in \cref{fig:room_simulation}.
The reverberation time is randomly sampled from [0.1, 0.4]~s.
The target sound source is randomly placed with a 0.5~m margin from the walls.
The close microphone is fixed 0.05~m from the target source, while the far microphone is randomly placed [0.1, 1.0]~m away.
Four different noise signals are randomly selected and played, each placed 0.2~m away from a corner of the room.
The SNR at the close microphone is randomly selected from [-10, 10]~dB.

\subsection{ASD procedure using NA-SSL models}
\Ac{nassl} models can be combined with well-established \ac{ssl}-based \ac{asd} techniques described in \cref{sec:ssl_asd}.
For the \ac{asd} backend, we simply use the estimated clean representation $\hat{\mathbf r}$ instead of the noisy representation.
Since the output of an \ac{nassl} model has the same format as that of the original \ac{ssl} model, any backend technique designed for \ac{ssl} representations can be directly applied to \ac{nassl} representations.

We can also perform discriminative fine-tuning of \ac{nassl} on top of $\hat{\mathbf r}$, where both the \ac{ssl} model and the NA layers are updated.
While the base \ac{ssl} model is shared between the close- and far-microphone signals, the NA layers refine only the close-microphone representation.
Therefore, fine-tuning the NA layers allows them to adapt specifically to the close-microphone representation, which predominantly contains the target machine sound.

\begin{table*}[t!]
    \centering
    \caption{
    Evaluation of frontends across three base \ac{ssl} models.
    The values are official scores, and Total shows the harmonic mean of the scores over the machine types.
    The mean, CI95, and ensemble results across the three trials are reported.
    The backend consists of BEAM and \ac{rdp} with $\gamma=4$.
    NA denotes \ac{nassl} pre-training, and Dis denotes discriminative fine-tuning.
    $^\ast$ denotes machine types without attribute labels.
    }
    \resizebox{\textwidth}{!}{
        \begin{tabular}{ll*{7}{c}cc}
            \toprule
            & & \multicolumn{8}{c}{Mean (CI95)} & \makecell[c]{Ensemble} \\
            \cmidrule(lr){3-10}\cmidrule(lr){11-11}
            Base & Method & bearingEmu$^\ast$ & fan & gearboxEmu & sliderEmu$^\ast$  & ToyCar$^\ast$  & ToyCarEmu & valveEmu$^\ast$  & Total & Total \\
            \midrule
            \multirow{4}{*}{BEATs} & Original & 63.27 {\scriptsize \makebox[2.5em][r]{}} & 52.73 {\scriptsize \makebox[3em][r]{}} & 64.63 {\scriptsize \makebox[2.5em][r]{}} & 64.42 {\scriptsize \makebox[2.5em][r]{}} & {\bfseries 58.85} {\scriptsize \makebox[2.5em][r]{}} & 58.70 {\scriptsize \makebox[2.5em][r]{}} & 69.68 {\scriptsize \makebox[3em][r]{}} & 61.32 {\scriptsize \makebox[2.5em][r]{}} & 61.32 \\
            & NA & 62.33 {\scriptsize \makebox[2.5em][r]{(4.77)}} & 52.19 {\scriptsize \makebox[3em][r]{(5.78)}} & 63.73 {\scriptsize \makebox[2.5em][r]{(1.50)}} & 69.59 {\scriptsize \makebox[2.5em][r]{(4.82)}} & 58.45 {\scriptsize \makebox[2.5em][r]{(1.52)}} & {\bfseries 59.27} {\scriptsize \makebox[2.5em][r]{(3.27)}} & 93.39 {\scriptsize \makebox[3em][r]{(3.97)}} & 63.64 {\scriptsize \makebox[2.5em][r]{(1.71)}} & 63.92 \\
            & Dis & 63.38 {\scriptsize \makebox[2.5em][r]{(0.33)}} & 51.47 {\scriptsize \makebox[3em][r]{(1.14)}} & 66.47 {\scriptsize \makebox[2.5em][r]{(1.38)}} & 66.43 {\scriptsize \makebox[2.5em][r]{(1.12)}} & 55.30 {\scriptsize \makebox[2.5em][r]{(2.17)}} & 57.15 {\scriptsize \makebox[2.5em][r]{(1.88)}} & 80.55 {\scriptsize \makebox[3em][r]{(1.17)}} & 61.78 {\scriptsize \makebox[2.5em][r]{(0.72)}} & 61.82 \\
            & Dis NA & {\bfseries 64.64} {\scriptsize \makebox[2.5em][r]{(7.99)}} & {\bfseries 53.47} {\scriptsize \makebox[3em][r]{(8.31)}} & {\bfseries 69.23} {\scriptsize \makebox[2.5em][r]{(1.32)}} & {\bfseries 72.00} {\scriptsize \makebox[2.5em][r]{(2.60)}} & 54.16 {\scriptsize \makebox[2.5em][r]{(4.13)}} & 57.77 {\scriptsize \makebox[2.5em][r]{(2.70)}} & {\bfseries 95.63} {\scriptsize \makebox[3em][r]{(0.94)}} & {\bfseries 64.32} {\scriptsize \makebox[2.5em][r]{(0.77)}} & {\bfseries 64.57} \\
            \midrule
            \multirow{4}{*}{EAT} & Original & {\bfseries 62.02} {\scriptsize \makebox[2.5em][r]{}} & 53.16 {\scriptsize \makebox[3em][r]{}} & 61.48 {\scriptsize \makebox[2.5em][r]{}} & 61.18 {\scriptsize \makebox[2.5em][r]{}} & 54.99 {\scriptsize \makebox[2.5em][r]{}} & {\bfseries 58.11} {\scriptsize \makebox[2.5em][r]{}} & 65.96 {\scriptsize \makebox[3em][r]{}} & 59.27 {\scriptsize \makebox[2.5em][r]{}} & 59.27 \\
            & NA & 61.21 {\scriptsize \makebox[2.5em][r]{(0.69)}} & 56.17 {\scriptsize \makebox[3em][r]{(4.80)}} & 64.06 {\scriptsize \makebox[2.5em][r]{(1.30)}} & 66.63 {\scriptsize \makebox[2.5em][r]{(6.97)}} & {\bfseries 55.40} {\scriptsize \makebox[2.5em][r]{(1.26)}} & 57.53 {\scriptsize \makebox[2.5em][r]{(0.17)}} & 88.11 {\scriptsize \makebox[3em][r]{(1.97)}} & 62.74 {\scriptsize \makebox[2.5em][r]{(1.41)}} & 63.23 \\
            & Dis & 61.63 {\scriptsize \makebox[2.5em][r]{(2.97)}} & 48.30 {\scriptsize \makebox[3em][r]{(10.03)}} & 62.61 {\scriptsize \makebox[2.5em][r]{(3.03)}} & 58.24 {\scriptsize \makebox[2.5em][r]{(0.90)}} & 50.45 {\scriptsize \makebox[2.5em][r]{(0.60)}} & 57.71 {\scriptsize \makebox[2.5em][r]{(1.36)}} & 66.54 {\scriptsize \makebox[3em][r]{(6.25)}} & 57.21 {\scriptsize \makebox[2.5em][r]{(2.59)}} & 56.72 \\
            & Dis NA & 61.56 {\scriptsize \makebox[2.5em][r]{(2.18)}} & {\bfseries 57.60} {\scriptsize \makebox[3em][r]{(2.58)}} & {\bfseries 68.73} {\scriptsize \makebox[2.5em][r]{(2.16)}} & {\bfseries 71.10} {\scriptsize \makebox[2.5em][r]{(3.46)}} & 54.27 {\scriptsize \makebox[2.5em][r]{(0.50)}} & 55.87 {\scriptsize \makebox[2.5em][r]{(1.75)}} & {\bfseries 90.80} {\scriptsize \makebox[3em][r]{(3.70)}} & {\bfseries 63.88} {\scriptsize \makebox[2.5em][r]{(0.30)}} & {\bfseries 64.20} \\
            \midrule
            \multirow{4}{*}{Dasheng} & Original & 64.36 {\scriptsize \makebox[2.5em][r]{}} & 52.58 {\scriptsize \makebox[3em][r]{}} & 59.97 {\scriptsize \makebox[2.5em][r]{}} & 57.75 {\scriptsize \makebox[2.5em][r]{}} & 52.36 {\scriptsize \makebox[2.5em][r]{}} & 55.26 {\scriptsize \makebox[2.5em][r]{}} & 58.47 {\scriptsize \makebox[3em][r]{}} & 56.98 {\scriptsize \makebox[2.5em][r]{}} & 56.98 \\
            & NA & 63.40 {\scriptsize \makebox[2.5em][r]{(2.14)}} & 53.92 {\scriptsize \makebox[3em][r]{(10.09)}} & 59.08 {\scriptsize \makebox[2.5em][r]{(8.00)}} & {\bfseries 65.65} {\scriptsize \makebox[2.5em][r]{(5.93)}} & {\bfseries 57.39} {\scriptsize \makebox[2.5em][r]{(1.74)}} & 54.58 {\scriptsize \makebox[2.5em][r]{(2.74)}} & 84.99 {\scriptsize \makebox[3em][r]{(13.63)}} & 61.33 {\scriptsize \makebox[2.5em][r]{(2.13)}} & 62.34 \\
            & Dis & 65.25 {\scriptsize \makebox[2.5em][r]{(2.61)}} & 52.95 {\scriptsize \makebox[3em][r]{(2.11)}} & {\bfseries 61.51} {\scriptsize \makebox[2.5em][r]{(1.02)}} & 57.49 {\scriptsize \makebox[2.5em][r]{(1.83)}} & 52.61 {\scriptsize \makebox[2.5em][r]{(0.55)}} & 54.35 {\scriptsize \makebox[2.5em][r]{(2.67)}} & 70.56 {\scriptsize \makebox[3em][r]{(3.39)}} & 58.60 {\scriptsize \makebox[2.5em][r]{(0.69)}} & 58.91 \\
            & Dis NA & {\bfseries 66.81} {\scriptsize \makebox[2.5em][r]{(3.51)}} & {\bfseries 54.81} {\scriptsize \makebox[3em][r]{(4.87)}} & 60.82 {\scriptsize \makebox[2.5em][r]{(1.24)}} & 64.87 {\scriptsize \makebox[2.5em][r]{(0.25)}} & 52.93 {\scriptsize \makebox[2.5em][r]{(1.46)}} & {\bfseries 55.71} {\scriptsize \makebox[2.5em][r]{(3.03)}} & {\bfseries 90.03} {\scriptsize \makebox[3em][r]{(2.65)}} & {\bfseries 61.93} {\scriptsize \makebox[2.5em][r]{(2.13)}} & {\bfseries 63.10} \\
            \bottomrule
            \end{tabular}
    }
    \label{tab:evaluation_machine_wise}
\end{table*}

\section{Experimental evaluation}
In the experimental evaluation, we demonstrate the general effectiveness of the \ac{nassl} framework across diverse settings, including three base \ac{ssl} models, multiple backend techniques, and configurations with and without discriminative fine-tuning.

\subsection{Setup}
We conducted experimental evaluations using the DCASE 2026 Challenge Task~2 dataset, compiled from ToyADMOS2~\cite{harada2021toyadmos2} and MIMII DG~\cite{dohi2022mimii}.
The dataset consists of development and evaluation sets with distinct machine types.
The development set includes seven machine types: bearingEmu, fan, gearboxEmu, valveEmu, sliderEmu, ToyCarEmu, and ToyCar.
The evaluation set includes five machine types: ToyDrone, ToothBrush, SewingMachine, Sander, and BlowerDustCollector.
Each machine type has 1000 training samples and 200 test samples.
All training samples are normal, with 99\% from the source domain and 1\% from the target domain.
Attribute labels are available only for the training data of the following machine types: fan, gearboxEmu, ToyCarEmu, ToyDrone, Sander, and BlowerDustCollector.
The test data in the development set consist of 50 samples for each combination of domain and normal/anomalous class.
The test data in the evaluation set are provided without domain and normal/anomalous labels for the official challenge evaluation.
Each recording is a 6--16-second two-channel signal sampled at 16~kHz.

For the NA layers, we set the number of heads in MHCA to 8 and the hidden size of the \ac{ffn} to 2304.
We trained the \ac{nassl} models for 200 epochs using the AdamW optimizer, a fixed learning rate of 0.0001, and a batch size of 160.
We applied \ac{ema} with a decay rate of 0.999 after 20k training steps.

For discriminative fine-tuning, we generated pseudo labels for machine types without attribute labels.
When fine-tuning a given \ac{ssl} or \ac{nassl} model, we generated the pseudo labels from that same model.
The pseudo labels were generated by applying k-means clustering to the frequency-preserving representation.
For each machine type, pseudo labels were generated in the source domain, where the number of clusters was set to 0.8\% of the number of samples.
For the target domain, all samples were assigned to a single target-domain class within each machine type.

During the discriminative fine-tuning, we aggregated the representation sequence $\mathbf r$ into a single 256-dimensional discriminative feature using an attentive statistics pooling layer~\cite{okabe2018interspeech} and a linear layer.
In the backend, the \ac{ssl} representation sequence before aggregation was used rather than the resulting discriminative feature.
For the fine-tuning of the original \ac{ssl} models, we applied \ac{lora} with a rank of 64 to the query, key, and value projection layers in the attention modules.
In \ac{nassl}, \ac{lora} was also applied to MHCA in the NA layers with the same configuration.
We fine-tuned the models for 25 epochs using sub-cluster AdaCos~\cite{wilkinghoff2021sub} with 16 sub-clusters, mixup with a probability of 0.5, AdamW optimizer, and a batch size of 8.
The learning rate was linearly increased from 0 to 0.0001 over the first 5,000 steps.

We evaluated the performance with multiple backend techniques: frequency-preserving representation, BEAM, and \ac{rdp}.
To address the data imbalance between the source and target domains, we always used variance-minimization-based score rescaling~\cite{matsumoto2025adjusting} with four nearest neighbors.
Cosine distance was used to calculate the anomaly score.

As the evaluation metric, we used the official score~\cite{nishida2026description} calculated as the harmonic mean of three types of \ac{auc}: (a) \ac{pauc} with $p=0.1$, calculated using samples from both the source and target domains; and (b) source-domain and (c) target-domain \acp{auc}, each calculated using normal samples from the corresponding domain and anomalous samples from both domains.
We performed \ac{nassl} pre-training and discriminative fine-tuning for three trials with different random seeds.
We report the mean and 95\% confidence interval (CI95) of the official scores across the three trials and also report the ensemble result obtained by averaging the anomaly scores across the trials.

\begin{table}[t!]
    \centering
    \caption{
    Evaluation using multiple backends.
    The values are harmonic mean of the official scores over machine types in the development dataset.
    The values are reported as ``Mean (CI95)''.
    NA denotes \ac{nassl} pre-training, and Dis denotes discriminative fine-tuning.
    Global and Freq denote global and frequency-preserving representations, respectively.
    AP denotes the average pooling.
    The value in parentheses after ``RDP'' indicates $\gamma$.
    Since Dasheng extracts segment-wise representations, the frequency-preserving representation is identical to the global representation.
    }
    \resizebox{\columnwidth}{!}{
    \begin{tabular}{rrllllll}
\toprule
 \makecell{\\ \\Base} & \makecell{\\ \\Method} & \makecell{Global\\AP\\\\} & \makecell{Freq\\AP\\\\} & \makecell{Freq\\AP\\+BEAM\\} & \makecell{Freq\\RDP (4)\\+BEAM}  & \makecell{Freq\\RDP (8)\\+BEAM} \\
\midrule

\multirow{4}{*}[-8pt]{BEATs} & Original & \makecell{57.16} & \makecell{57.94} & \makecell{60.28} & \makecell{61.32} & \makecell{{\bfseries 62.02}} \\
& NA & \makecell{60.15\\[-2.8pt]{\scriptsize (1.08)}} & \makecell{60.25\\[-2.8pt]{\scriptsize (2.74)}} & \makecell{62.47\\[-2.8pt]{\scriptsize (1.88)}} & \makecell{63.64\\[-2.8pt]{\scriptsize (1.71)}} & \makecell{{\bfseries 64.10}\\[-2.8pt]{\scriptsize (1.81)}} \\
& Dis & \makecell{59.66\\[-2.8pt]{\scriptsize (0.66)}} & \makecell{58.53\\[-2.8pt]{\scriptsize (0.72)}} & \makecell{61.41\\[-2.8pt]{\scriptsize (0.52)}} & \makecell{61.78\\[-2.8pt]{\scriptsize (0.72)}} & \makecell{{\bfseries 61.91}\\[-2.8pt]{\scriptsize (0.98)}} \\
& Dis NA & \makecell{60.96\\[-2.8pt]{\scriptsize (2.42)}} & \makecell{61.57\\[-2.8pt]{\scriptsize (0.98)}} & \makecell{63.52\\[-2.8pt]{\scriptsize (1.39)}} & \makecell{64.32\\[-2.8pt]{\scriptsize (0.77)}} & \makecell{{\bfseries 64.35}\\[-2.8pt]{\scriptsize (1.44)}} \\
\midrule
\multirow{4}{*}[-8pt]{EAT} & Original & \makecell{55.56} & \makecell{56.61} & \makecell{59.24} & \makecell{{\bfseries 59.27}} & \makecell{57.49} \\
& NA & \makecell{58.93\\[-2.8pt]{\scriptsize (0.75)}} & \makecell{60.62\\[-2.8pt]{\scriptsize (1.01)}} & \makecell{62.59\\[-2.8pt]{\scriptsize (1.50)}} & \makecell{{\bfseries 62.74}\\[-2.8pt]{\scriptsize (1.41)}} & \makecell{62.42\\[-2.8pt]{\scriptsize (1.32)}} \\
& Dis & \makecell{{\bfseries 59.21}\\[-2.8pt]{\scriptsize (1.79)}} & \makecell{56.99\\[-2.8pt]{\scriptsize (0.25)}} & \makecell{58.69\\[-2.8pt]{\scriptsize (1.91)}} & \makecell{57.21\\[-2.8pt]{\scriptsize (2.59)}} & \makecell{55.88\\[-2.8pt]{\scriptsize (1.35)}} \\
& Dis NA & \makecell{62.80\\[-2.8pt]{\scriptsize (1.46)}} & \makecell{62.69\\[-2.8pt]{\scriptsize (0.63)}} & \makecell{{\bfseries 63.99}\\[-2.8pt]{\scriptsize (0.32)}} & \makecell{63.88\\[-2.8pt]{\scriptsize (0.30)}} & \makecell{63.12\\[-2.8pt]{\scriptsize (2.36)}} \\
\midrule
\multirow{4}{*}[-8pt]{Dasheng} & Original & \makecell{56.66} & \makecell{-} & \makecell{-} & \makecell{56.98} & \makecell{{\bfseries 57.07}} \\
& NA & \makecell{61.28\\[-2.8pt]{\scriptsize (3.71)}} & \makecell{-} & \makecell{-} & \makecell{61.33\\[-2.8pt]{\scriptsize (2.13)}} & \makecell{{\bfseries 61.75}\\[-2.8pt]{\scriptsize (0.64)}} \\
& Dis & \makecell{58.33\\[-2.8pt]{\scriptsize (0.42)}} & \makecell{-} & \makecell{-} & \makecell{58.60\\[-2.8pt]{\scriptsize (0.69)}} & \makecell{{\bfseries 59.07}\\[-2.8pt]{\scriptsize (0.92)}} \\
& Dis NA & \makecell{61.72\\[-2.8pt]{\scriptsize (2.53)}} & \makecell{-} & \makecell{-} & \makecell{61.93\\[-2.8pt]{\scriptsize (2.13)}} & \makecell{{\bfseries 62.63}\\[-2.8pt]{\scriptsize (1.89)}} \\

\bottomrule
\end{tabular}
}
\label{tab:evaluation_backend}
\end{table}

\subsection{Results}

\cref{tab:evaluation_machine_wise} compares frontends using the same backend consisting of BEAM and \ac{rdp} with $\gamma=4$.
First, we can see that \ac{nassl} is effective regardless of the base model and whether discriminative fine-tuning is applied.
\ac{nassl} yields substantial improvements of over 20\% for valveEmu and also improves performance for the other machine types.
Discriminative fine-tuning further improves performance and achieves the best performance for all base models.
Also, by using pseudo labels, the benefits of discriminative fine-tuning can be observed even for machine types without attribute labels.
We also observe high variance in \ac{asd} performance for \ac{nassl} models, highlighting the importance of ensembling multiple trials.

\Cref{tab:evaluation_backend} shows the results obtained using different backend techniques.
The results demonstrate the effectiveness of the individual backend techniques and show that their benefits also carry over to \ac{nassl}-based frontends.
In particular, we can see that BEAM substantially improves performance.
For \ac{rdp}, larger values of $\gamma$ yield greater improvements for BEATs and Dasheng, whereas \ac{rdp} provides little benefit for EAT. These trends observed for \ac{rdp} are consistent with the findings reported in~\cite{wilkinghoff2026temporal}.

We also conducted an ablation study to evaluate the effect of applying \ac{lora} to the NA layers during discriminative fine-tuning, as shown in \cref{tab:evaluation_lora}.
Although discriminative fine-tuning remains effective even without \ac{lora} in the NA layers, applying \ac{lora} to the NA layers tends to yield better performance.

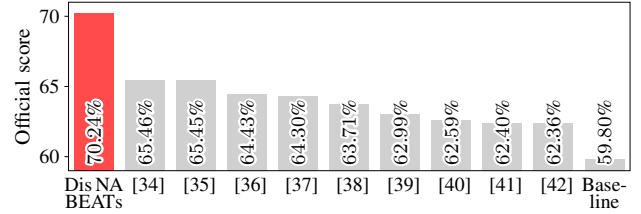
\begin{figure}[t]
    \centering
    \begin{adjustbox}{max width=\columnwidth}
    \definecolor{scorebarred}{HTML}{FF4B4B}
\definecolor{scorebargray}{HTML}{D0D0D0}
\contourlength{1pt}

\begin{tikzpicture}[
    scale=0.9,
    x=0.8cm,
    y=0.22cm,
    axis/.style={draw=black, line width=0.35pt},
    ticklabel/.style={font=\footnotesize, inner sep=1pt},
    xlabel/.style={font=\footnotesize, inner sep=1pt, align=center},
    scorelabel/.style={font=\footnotesize, rotate=90, anchor=west, inner sep=0pt}
]

\def\ymin{59}
\def\ymax{71}
\def\xmin{-0.5}
\def\xmax{10.5}

\def\scorebarentry#1#2#3#4{%
    \path[fill=#4, draw=none]
        ({#1-0.39},0) rectangle ({#1+0.39},{#2-\ymin});
    \node[xlabel, anchor=north] at (#1,-0.25) {#3};
    \node[scorelabel] at (#1,0.4) {\contour{white}{\pgfmathprintnumber[fixed zerofill, precision=2]{#2}\%}};
}

\scorebarentry{0}{70.2412}{Dis\,NA\\[-2pt]BEATs}{scorebarred}
\scorebarentry{1}{65.4621}{\cite{wu2026unsupervised}}{scorebargray}
\scorebarentry{2}{65.4529}{\cite{yu2026noise}}{scorebargray}
\scorebarentry{3}{64.4277}{\cite{qian2026anomalous}}{scorebargray}
\scorebarentry{4}{64.2960}{\cite{yang2026dual}}{scorebargray}
\scorebarentry{5}{63.7101}{\cite{zhang2026thuee}}{scorebargray}
\scorebarentry{6}{62.9941}{\cite{fan2026wistlab}}{scorebargray}
\scorebarentry{7}{62.5878}{\cite{krag2026online}}{scorebargray}
\scorebarentry{8}{62.4013}{\cite{zhou2026eat}}{scorebargray}
\scorebarentry{9}{62.3631}{\cite{jiang2026aithu}}{scorebargray}
\scorebarentry{10}{59.8029}{Base-\\[-2pt]line}{scorebargray}

\draw[axis] (\xmin,0) rectangle (\xmax,{\ymax-\ymin});

\foreach \ytick in {60,65,70} {
    \draw[axis]
        (\xmin,{\ytick-\ymin})
        -- ({\xmin-0.08},{\ytick-\ymin});

    \node[ticklabel, anchor=east]
        at ({\xmin-0.12},{\ytick-\ymin}) {\ytick};
}

\node[font=\small, rotate=90, anchor=south]
    at ({\xmin-0.6},{(\ymax-\ymin)/2}) {Official score};

\def\xmargin{1.4}

\pgfresetboundingbox
\path[use as bounding box]
    ({\xmin-1.0},-2)
    rectangle
    ({\xmax+0.5},{\ymax-\ymin+0.3});
    
\end{tikzpicture}
    \end{adjustbox}
    \caption{
    Comparison of our Dis NA-BEATs with the ten top-performing systems and the official baseline system on the DCASE 2026 Challenge Task~2 evaluation dataset.
    }
    \label{fig:score_bar}
\end{figure}

\Cref{fig:score_bar} shows the results on the DCASE 2026 Challenge Task~2 evaluation dataset reported in \cite{DCASE2026Task2Homepage}.
It compares our Dis NA-BEATs with the ten top-performing systems~\cite{wu2026unsupervised,yu2026noise,qian2026anomalous,yang2026dual,zhang2026thuee,fan2026wistlab,krag2026online,zhou2026eat,jiang2026aithu} and the official baseline system~\cite{nishida2026description}. %
Here, the backend consists of BEAM and \ac{rdp} with $\gamma=4$, and the anomaly scores from three trials are ensembled.
Dis NA-BEATs ranked first on the evaluation dataset with an official score of 70.24\%, surpassing the second-place system (65.46\%) and the official baseline system (59.80\%) by a considerable margin.
It is also noteworthy that our Dis NA-BEATs is built on only a single base \ac{ssl} model with a single backend, whereas the other systems rely on ensembles of multiple \ac{ssl} models~\cite{yu2026noise,krag2026online} and multiple backends~\cite{yu2026noise,zhang2026thuee,fan2026wistlab,krag2026online,jiang2026aithu}.
Moreover, our Dis NA-BEATs outperforms the other BEATs-based systems~\cite{yu2026noise,fan2026wistlab,krag2026online,jiang2026aithu}.

\begin{table}[t!]
    \centering
    \caption{Ablation study of discriminative fine-tuning.
    The values are harmonic mean of the official scores over machine types in the development dataset.
    The values are reported as ``Mean (CI95)''.
    The backend consists of BEAM and \ac{rdp} with $\gamma=4$.
    NA denotes \ac{nassl} pre-training, and Dis denotes discriminative fine-tuning.
    }
    \resizebox{\columnwidth}{!}{
    \begin{tabular}{lrrr}
    \toprule
    Method & BEATs & EAT & Dasheng \\
    \midrule
    NA & 63.64 {\scriptsize (1.71)} & 62.74 {\scriptsize (1.41)} & 61.33 {\scriptsize (2.13)} \\
    Dis NA  & {\bfseries 64.32} {\scriptsize (0.77)} & {\bfseries 63.88} {\scriptsize (0.30)} & 61.93 {\scriptsize (2.13)} \\
    Dis NA w/o LoRA in NA layers & 63.97 {\scriptsize (1.81)} & 63.17 {\scriptsize (1.17)} & {\bfseries 62.09} {\scriptsize (0.84)} \\
    \bottomrule
    \end{tabular}
    }
    \label{tab:evaluation_lora}
\end{table}

\section{Conclusion}
In this paper, we explored the use of \ac{nassl} models for the \ac{naasd} task.
The \ac{nassl} models aim to extract clean \ac{ssl} representations from noisy signals by leveraging auxiliary noise information captured by a microphone located farther from the target sound source.
These models are first pre-trained on simulated two-channel recordings constructed from diverse audio signals.
They are then used as frontends within a well-established \ac{ssl}-based \ac{asd} framework.
Our experimental results demonstrated the general effectiveness and broad compatibility of the \ac{nassl} framework across three base \ac{ssl} models, multiple backend methods, and configurations both with and without discriminative fine-tuning.

\bibliographystyle{IEEEtran}
\bibliography{refs}

\end{document}